\documentclass{aa}
\usepackage{txfonts}
\usepackage{natbib}
\usepackage {float}
\usepackage{graphicx}
\usepackage{xcolor}

\newcommand{\bq}{\begin{equation}} 
\newcommand{\eq}{\end{equation}}   

\newcommand{\beq}{\begin{equation}}
\newcommand{\eeq}{\end{equation}}
\newcommand{\beqa}{\begin{eqnarray}}
\newcommand{\eeqa}{\end{eqnarray}}

\newcommand{\PL}{$P\hbox{--}L$}

\newcommand{\muas}{\hbox{$\, \mu\rm as$}}

\newcommand{\mas}{\hbox{$\, \rm mas$}}
\newcommand{\kpc}{\hbox{$\, \rm kpc$}}
\newcommand{\lcdm}{\hbox{$\Lambda$CDM}}

\newcommand{\hunit}{\hbox{$\rm km \, s^{-1} \, Mpc^{-1} $}}

\newcommand{\numgoodcephs}{202}

\long\def\hide#1{}
\long\def\newtext#1{{#1}}

\newcommand{\medianTGASfractionalerror}{\hbox{50\%}}
\newcommand{\tgasmultfactor}{\hbox{0.997}}
\newcommand{\tgasmulterror}{\hbox{0.025}}
\newcommand{\tgasmultfactorunity}{\hbox{0.997}}
\newcommand{\tgasmulterrorunity}{\hbox{0.031}}

\def\HST{{\it HST}}

\begin{document} 

\title{A test of Gaia Data Release 1 parallaxes: implications for the local distance scale}


\author {Stefano Casertano \inst{1,2} \and Adam G. Riess \inst{2,1} \and Beatrice Bucciarelli \inst{3} \and Mario G.~Lattanzi \inst{3}}

\institute { Space Telescope Science Institute, 3700 San Martin
  Drive, Baltimore, MD 21218, USA
   \and
  Department of Physics and Astronomy, Johns Hopkins
  University, Baltimore, MD 21218, USA
   \and
  INAF, Osservatorio Astronomico di Torino, Strada Osservatorio 20, 
     I-10025 Pino Torinese, TO, Italy
}

\abstract {}{
  We present a comparison of Gaia Data Release 1 (DR1) parallaxes with
  photometric parallaxes for a sample of 212 Galactic Cepheids at a
  median distance of 2~kpc, and explore their implications on the
  distance scale and the local value of the Hubble constant $ H_0 $.
}{
  The Cepheid distances are estimated from a recent calibration of the
  near-infrared Period-Luminosity (\PL) relation.
  The comparison is carried out in parallax space, where the DR1 parallax
  errors, with a median value of half the median parallax, are expected
  to be well-behaved.  
}{
  With the exception of one outlier, the DR1 parallaxes are in
  remarkably good global agreement with the predictions, and there is
  an indication that the published errors may be conservatively
  overestimated by about 20\%.  Our analysis suggests that the
  parallaxes of 9 Cepheids brighter than $G = 6$ may be systematically
  underestimated; {\it trigonometric} parallaxes measured with the
  Hubble Space Telescope Fine Guidance Sensor for three of these
  objects confirm this trend.  If interpreted as an independent
  calibration of the Cepheid luminosities and assumed to be otherwise
  free of systematic uncertainties, DR1 parallaxes would imply a
  decrease of 0.3\% in the current estimate of the local Hubble
  constant, well within their statistical uncertainty, and
  corresponding to a value 2.5 $\sigma$ (3.5 $\sigma$ if the errors
  are scaled) higher than the value inferred from Planck CMB data used
  in conjunction with {\lcdm}.  We also test for a zeropoint error in
  Gaia parallaxes and find none to a precision of $ \sim 20 \muas$.
  We caution however that with this early release, the complete
  systematic properties of the measurements may not be fully
  understood at the statistical level of the Cepheid sample mean, a
  level an order of magnitude below the individual uncertainties.  The
  early results from DR1 demonstrate again the enormous impact that
  the full mission will likely have on fundamental questions in
  astrophysics and cosmology.
}{}



\maketitle

\section {Introduction}

The Gaia Mission \citep{prusti12, prusti16}, launched in December 2013
by the European Space Agency, will revolutionize our knowledge of
individual stellar objects and of the structure of the Milky Way by
providing distance and velocity measurements of unprecedented
precision for over a billion individual objects within the Milky Way.
Gaia end-of-mission results will also have a major impact on
cosmology, enabling the determination of the Hubble Constant $ H_0 $
to better than $ 1\% $, assuming concomitant control of statistical and
systematic errors beyond the parallax measurements \citep[hereafter
R16]{riess16}.  The mission is expected to last at least 5 years, and
the final astrometric measurements, with expected accuracy $ \sim
5\hbox{--}10 \muas $ for the best-measured stars, will be released in
2022.

The Gaia DR1 \citep{brown16}, which took place September 14, 2016, is
based on measurements taken within the first 14 months of
observations \citep{lindegren16}, and includes data for
over a billion stars brighter than magnitude 20.7.  Due to the short
observing period and the limited number of separate measurements for
each target, parallax and proper motion are partially degenerate, and
for a majority of the star only positions are available in DR1.
However, for about 2 million stars in common with the Hipparcos
\citep{perryman97, vanleeuwen07a} and Tycho-2 \citep{hoeg00} catalogs,
parallaxes and proper motions could be determined as part of the
Tycho-Gaia Astrometric Solution (hereafter TGAS; \citealt{michalik15})
and are included in DR1.  These stars are typically brighter than
visual magnitude 11.5 and have typical parallax uncertainties of $ 300
\muas $.

One of the early difficulties encountered by the Gaia mission has been
a somewhat unexpected variation of the Basic Angle, the angular
separation between the two fields of view, with an amplitude of
several mas, as reported by the on-board Basic Angle Monitor (BAM).
Data taken during commissioning \citep{mora14} prove that the
variation reported by the BAM is real.  Self-calibration procedures
have been devised to correct for the Basic Angle variation, and it is
expected that the residuals are degenerate with a global zero
point error in the parallax \citep{michalik16}.  A sample of 135,000
AGN has been used as part of the solution and its verification; from
an analysis of parallaxes for both Hipparcos stars and AGN,
\citet{lindegren16} suggest that parallax zero point errors are $
\lesssim 0.1 \mas $ in magnitude, with possible zonal variations and
color terms, and a systematic difference between Northern and Southern
hemispheres.

\citet{lindegren16} offer several observational and scientific tests
and validations of the TGAS results.  Because of the inherent
challenge in reaching a new level of parallax precision, it is
important to produce additional tests of the set of Gaia parallax
measurements. Here we employ an additional validation, based on the
sample of 249 Galactic Cepheids in \citet{vanleeuwen07b}, at a median
distance of 2 kpc.  A similar test, based on the sample of Cepheids in
\citet{fouque07} but excluding nearby Cepheids (within 1 kpc),
is reported in Section~C3 of \citet{lindegren16}.

The Cepheid {\PL} relation, called the Leavitt Law \citep{leavitt12},
provides a tight correlation between period and luminosity of
Fundamental Mode Cepheids and has been central to the determination of
the scale of the Universe for a century.  Yet most Galactic Cepheids
are at a distance of a few kpc; until recently, this placed them well
beyond the useful range for measurements of accurate trigonometric
parallaxes.  \citet{benedict02, benedict07} used the Fine Guidance
Sensors (FGS) on the Hubble Space Telescope (\HST) to measure the
parallaxes of ten Cepheids within $ 0.5 $~\kpc, reaching individual
precisions of 150 to 300 {\muas}, or an average of 8\% per object.\
More recently, \citet{riess14} and \citet{casertano16} measured two
Cepheid parallaxes using {\HST} Wide Field Camera 3 in scanning mode,
with precision of 54 and 38 {\muas}, respectively.  With the addition
of a few Hipparcos measurements with errors of $ 300 \muas $, the
resulting sample of 15 Galactic Cepheids had a weighted mean precision
of 2.1\% and provided one of three anchors for the determination of
the Hubble constant \citep[R16; see also][]{riess11, freedman12}.  As
discussed in Section~2, this precision requires the use of
reddening-insensitive Wesenheit magnitudes \citet{madore82} anchored
in the near-infrared (R16, \citealt{macri15}), and a careful
calibration of the characteristics of the {\PL} relation.

In principle, TGAS parallaxes could provide a test of the zero point
of the {\PL} relation and thus contribute an independent calibration
of the local distance scale (see Section~4); however, given the early
nature of this release, we caution that systematic effects an order of
magnitude below the typical measurement error, yet large enough to
affect the resulting calibration, are difficult to rule out, thus the
results of this analysis must be seen as tentative.

The rest of this Letter is organized as follows.  In Section 2 we
describe the Cepheid sample and the underlying data on which the
comparison is based.  In Section 3 we carry out the test of the
quality of TGAS parallaxes for stars in our sample.  In Section 4 we
provide a tentative analysis of the implications for the zero point
calibration of the Leavitt Law and for the determination of the Hubble
constant $ H_0 $.  Finally, in Section 5 we summarize our conclusion
and indicate the potential impact of the full-mission Gaia results on
the determination of $ H_0 $.

\section {Data collected for the Cepheid parallax test}

The main sample we use to validate and interpret the Gaia TGAS
parallax measurements is a set of 249 Milky Way Cepheids identified by
\citet{vanleeuwen07b} as having Hipparcos measurements, together with
the ground-based optical and near-infrared photometry included
therein.  We estimate the distance of these Cepheids on the basis of
the infrared {\PL} relation as calibrated by R16.  In order to reduce
the impact of object-by-object reddening and extinction, the {\PL}
relation is often formulated in terms of a so-called Wesenheit
magnitude \citep[see R16 for details of our implementation]{madore82,
macri15}.  Wesenheit magnitudes are formed by subtracting from the
primary magnitude a color term in the same direction as the reddening
law; if the spectra are smooth and the reddening law is well
constrained, a Wesenheit magnitude is then reddening-free, in the
sense that a given source will have the same Wesenheit magnitude when
exposed to different degrees of reddening.  For Cepheids, Wesenheit
magnitudes, especially those for which the primary filter is in the
near-IR, have the additional advantages that the color term is small
and insensitive to the reddening raw, the {\PL} relation is
insensitive to metallicity, and its intrinsic width is reduced.
We use the same quantity as in R16:
 \beq
 m_H^W = m_{160} - 0.3861*(m_{555}-m_{814})
 \eeq
where $ m_{160} $, $ m_{555} $, and $ m_{814} $ are the Vega-system
magnitudes in the WFC3 filters F160W (near-IR), F555W, and F814W, with
central wavelength 1537, 531, and 802 nm, respectively.  The
coefficient $ 0.3861 $ is appropriate to a Galactic reddening law
\citep{fitzpatrick99} with $ R = 3.3 $.  
\hide {In this magnitude system, the {\PL} relation has an intrinsic
scatter of $ 0.08 $~mag \citep{macri15}.}
Since the available photometry in \citet{vanleeuwen07b} is
ground-based, while the best calibration of the {\PL} relation
obtained in R16 is based on {\HST} data, we use the ground-to-{\HST}
transformations given in R16, Equations~10--12.  The systematic
uncertainty in this transformation, found by R16 to be 0.013 mag, is
small compared to the 0.05 mag precision for the mean of the sample
(see Section~3).  However, to retain the full precision of {\it
future} Gaia Cepheid parallax measurements it will be crucial to
measure the Milky Way Cepheid mean magnitudes directly with {\HST},
as discussed in Section~5.

We adopt the primary calibration of the {\PL} relation from R16:
 \beq
 M_H^W = -2.77  - 3.26 * \log(P) 
 \eeq
where $ M_H^W $ is the absolute magnitude in the same Wesenheit system
of Equation~1, and $ P $ is the Cepheid period in days.  The slope of
this relation is constrained to 0.02 (0.6\%) by 2500 Cepheids measured
in over 20 different systems, including the Large Magellanic Cloud,
M31, NGC~4258, and galaxies that hosted a Type Ia supernova.  These
Cepheids also demonstrate the lack of any significant break in slope
near $P=10$ days which appears in optical bands \citep{ngeow05a,
ngeow05b}.

The zero point of the {\PL} relation was determined by R16 by using
three independent geometric calibrations: trigonometric parallaxes for
the 15 Galactic Cepheids mentioned in the Introduction, eight
late-type detached eclipsing binaries in the Large Magellanic Cloud
\citep{pietrzynski13}, and the geometric distance measurement based on
the kinematic of a ring of OH masers in the galaxy NGC~4258
\citep{humphreys13}.  Together, these measurements provide a
calibration of the intercept in Equation~2 with an uncertainty of
0.035 mag (1.6\% in distance).  This calibration relates directly to
the value of $ H_0 $; its role in our comparison will be discussed
further in Section~3.

The absolute magnitude of each Cepheid from Equation~2, combined with
its reddening-free apparent magnitude from Equation~1, together
provide a photometric parallax estimate for each individual Cepheid.
The main statistical error in the photometric parallaxes is due to
the intrinsic width of the {\PL} relation, which for $ M_H^W $ is
estimated to be $ 0.08 $~mag in the LMC \citep{macri15}; errors due to
photometric measurements are much smaller in comparison.  Therefore
the statistical uncertainty in each parallax estimate is $ \sim 4\% $---substantially
smaller than the median measurement error of
{\medianTGASfractionalerror} reported for the TGAS solution (see
Section~3)---thus providing a suitable test of the quality of the TGAS
parallaxes.  However, these uncertainties will be very significant
when the full mission results are available and the typical Gaia
parallax uncertainty per star will likely be $ < 3\% $.
Systematic uncertainties, as in nearly all distance scale
projections, will result primarily in a {\it multiplicative} scaling
factor on the photometric parallax estimates, discussed 
further in Sections~3 and~4.

\section {Parallax comparison}

As described in Section~2, the available magnitude and period
information allows us to predict the parallax for the 212 stars in the
\citet{vanleeuwen07b} list for which photometry is available in the V,
I, J, and H band and for which a TGAS parallax is available in DR1.

We carry out our analysis in parallax space, rather than in distance
or photometry, because of the low signal-to-noise ratio (SNR, median
value $\sim 2 $) of individual parallaxes.  The determination of the
absolute magnitude of a Cepheid follows from its apparent magnitude
and parallax, where $M=m\!-\!\mu_\pi$ and $\mu_\pi$ is the distance
modulus derived from parallaxes, including standard corrections for
bias (often referred to as Lutz-Kelker bias) arising from the finite
SNR of parallax measurements \citep{lutz73, hanson79}.  For SNR $ < 10
$, the bias correction becomes large and complex, as one must contend
with a skewing of the likelihood due to the selection bias (of
presumably a disk population) which is asymmetric with distance and a
non-Gaussian conversion from parallax to magnitudes.  A number of
papers have argued about the optimal way to contend with these issues
for low SNR parallax measurements \citep{francis14, feast02, hanson79,
benedict07}.  The approach we have adopted here is to use the high
precision of Cepheid magnitudes and {\PL} parameters (SNR $\sim$ 100)
in R16 to predict the parallaxes to the Gaia DR1 parallaxes and
compare the results.  As a result, we are using the Gaia first release
catalogue as a test of the R16 results, rather than as an independent
calibration of the Cepheid {\PL}.  We expect the best use of the Gaia
results to change dramatically as the Gaia precision improves by more
than an order of magnitude.

The distribution of TGAS vs.~photometric parallaxes for the 212
Cepheids is shown in Figure~1, with the error bars indicating the
reported TGAS uncertainty.  Clearly the distributions are in good
basic agreement, with the exception of the outlier RW~Cam (labeled in
Fig.~1), which has a TGAS parallax of $ 3.687 \pm 0.797 $ mas, vs.~a
predicted value of 0.608 mas.  RW~Cam is known to have a very
luminous, B8.2~III companion \citep{evans94}, instead of the typical
main-sequence companion \citep{bohm-vitense85}; \citet{fernie00}
concludes that the bright companion is the cause of its unusual
photometric properties.  The bright companion can affect both the
photometric distance determination and the parallax measurement (see,
e.g., the discussion in \citealt{anderson16}).  Therefore we exclude
RW~Cam from further analysis.

\hide{The presence of $ \sim 1\% $ outliers is not surprising in such
early results; aside from anomalous measurement errors, astrophysical
causes, e.g., binarity \citep {riess14, casertano16, anderson16b}, can
easily cause anomalous parallax measurements, especially over such a
short observation period.  We caution therefore against drawing
conclusions intolerant to a $ \sim 1\% $ outlier rate from other DR1
parallax samples.}

With the exception of this outlier, most points in Figure~1 lie within
the nominal 1-$\sigma$ error bars, suggesting that perhaps the errors
may be overestimated.  Indeed, under the simplistic model that the
photometric and TGAS parallax are equal aside from measurement errors,
the $ \chi^2 $ per degree of freedom (no free parameters) is 0.63 (a
less than 1 in $10^5$ chance).  If we assume that the errors are
underestimated by a constant multiplicative factor, its likely value
is $1-\sqrt{0.63} $, or about 20\%.  While the assumption of a single
multiplicative factor may be simplistic, the conclusion that the error
are underestimated---at least for the sample of Cepheids for which we
carry out the comparison---appears solid, as there is no commonality
of information between our photometric parallax estimates and the TGAS
parallax measurements.  \citet{lindegren16} indicate that the formal
uncertainties reported by the solution process have been inflated by a
factor $ F $ derived from a comparison with Hipparcos parallaxes (see
their Equation~4 and Appendix B; $ F $ has a minimum value of 1.4).
Our comparison suggests that, at least for the stars we consider, this
inflation factor may have been overestimated.  In the following we
will consider two options: the errors as reported, and scaled errors
that are 0.8 times the values reported.

In addition to the inflation factor applied to the formal errors,
\citet{lindegren16} also indicate that there may be an additional
systematic error of $ 0.3 \mas $ on the reported parallaxes, and that
parallax errors may be correlated on scales up to $ 10^\circ $.  We
tested for the possibility of a correlation by analyzing the residuals
with respect to the photometric parallax estimates.  The Cepheids
fainter than $ G = 6 $~mag in our sample form 20301 unique pairs, of
which 0.5\% are separated by $ < 2^\circ $, and 6.7\% by $ < 10^\circ
$.  (Parallaxes for brighter stars may be systematically
underestimated, as discussed below.)  Therefore our Cepheids are very weakly 
correlated according to the \citet{lindegren16} criteria.
These relatively close pairs offer the opportunity to estimate the
angular correlation of parallaxes directly from the data.  We find
that the two-point correlation between residuals in the 1363 unique
pairs separated by $ < 10^\circ $ is $ 19 \pm 34 \muas $, suggesting
that the correlation on such scales is too small to be detected with
this test.  Note that our test treats all separations smaller than $
10^\circ $ equally, and is therefore insensitive to a correlation that
changes sign within that range (e.g., as suggested by Figure~D3 in
\citealt {lindegren16}).  Nonetheless, our finding of {\it smaller}
errors than reported suggests that additional systematics and
correlations, if they exist, are significantly smaller than the formal
uncertainties.

The distribution shown in Figure~1 also suggests that for some of the
largest photometric parallaxes ($ > 1.7 \mas $), the TGAS parallaxes
may be consistently low.  These stars are also among the brightest in
the sample.  All of the Cepheids in our sample are brighter than the
Gaia saturation limit ($ \sim 12 $~mag in the Gaia $ G $ band) and
therefore require the use of gating \citep{prusti16, fabricius16,
lindegren16, brown16} to be measured; the details of the process
depend on the brightness of each individual star.  In Figure~2 we show
the ratio of measured to predicted parallax as a function of average $
G $ magnitude as measured by Gaia; the red symbols show the average
ratio in bins 2 mag wide.  For stars fainter than $ G = 6 $ mag, the
unweighted average ratio between TGAS and predicted parallax is $ 1.03
\pm 0.04 $ (uncertainty in the mean); for the 9 stars brighter than $
G = 6 $ mag, the average ratio is $ 0.86 \pm 0.06 $.  A formal
significance test indicates a significant difference but is of limited
value, as the error distribution is not fully characterized (more on
this below); however, stars brighter than $ G \sim 6 $ are in fact
treated slightly differently from fainter stars.  At $ G = 6 $, a
larger download window is used \citep{fabricius16}; near this
magnitude, the core of the stellar image will likely saturate at the
shortest gating interval used (TDI gate 4, 16 lines;
\citealt{prusti16}).  Thus a separation between stars brighter and
fainter than $ G \sim 6 $~mag is naturally driven by the specifics of
the measurement process and merits the {\it a priori} check.
Additional confirmation that the difference is likely in the TGAS
measurements can be obtained by considering the 3 stars for which
$G<6$ and an independent trigonometric parallax measurement has been
obtained with {\HST} \citep {benedict02, benedict07, vanleeuwen07b}.  For
these, the Gaia DR1 values are also low, with a mean which is 0.70
$\pm$ 0.35 mas smaller than the independent parallaxes, indicating
that the apparent difference between TGAS and photometric parallaxes
for $ G < 6 $ is {\it not} due to any issues with photometric parallax
estimates for nearby, bright stars.  To be conservative, we exclude
the 9 stars brighter than $ G=6 $~mag from further analysis; their
exclusion does not change significantly the value of the reduced $
\chi^2 $ or the apparent error overestimate.

There are thus {\numgoodcephs} Cepheids from \citet{vanleeuwen07b}
with parallaxes in the TGAS sample that provide the means to explore
the present distance scale in R16 or alternatively the possibility of
a zeropoint parallax error in Gaia.  Despite the one outlier and the
possible underestimation of parallaxes for the few Cepheids brighter
than $ G = 6 $~mag, the agreement between the TGAS parallaxes and the
photometric distances predicted from the R16 Cepheid calibration is
remarkably good.  This is consistent with the comparison in
\citet{lindegren16}, who use 141 stars in the \citet{fouque07} sample,
likewise excluding the nearest Cepheids (within 1~kpc).

In Figure~3 we show the change in $ \chi^2 $ that would result from
either a multiplicative term in the photometric parallaxes (left
panel) or an additive term in the TGAS parallaxes (right panel).  The
formal minimum for the multiplicative term is obtained for a value of
$ \tgasmultfactor \pm \tgasmulterror $ with 0.8 scaled errors and $
\tgasmultfactorunity \pm \tgasmulterrorunity $ for nominal errors,
consistent with no difference.  For a global additive term, we find a
preferred value of $ 1 \muas $, with a formal uncertainty of $ 15
\muas $ for scaled errors ($ 19 \muas $ for nominal errors.  An
additive term could arise, e.g., from a parallax zero point error due
to from imperfections in the correction for Basic Angle variations
\citep{michalik16}.  \citet{lindegren16} carry out several tests for
zero point errors, finding terms with magnitude $ \sim 0.1 \mas $ and
possible dependence on hemisphere and color; given the small number of
sources in our sample, we limit our analysis to a single global zero
point offset, which we find to be less than $ 20 \muas $ (1-$\sigma$).

Note that in our analysis, the assumption is implicitly made that the
term in consideration (additive or multiplicative) is the {\it only}
systematic in the data.  Additional systematics could exist (e.g.,
terms that depend on magnitude, color, phase of observations, or
location in the sky), and if they are below the measurement errors,
would not be possible to detect without a specific test, but could
reduce the precision with which an additive or multiplicative term can
be determined.

We conclude that the TGAS parallaxes for the sample of Galactic
Cepheids under consideration are in very good agreement, to the limit
of their current precision, with the predicted parallaxes based on the
calibration of R16.  There is an indication and independent
confirmation of an underestimation of parallax for the few stars
brighter than $ G = 6 $ mag, and we estimate that the reported errors
appear conservatively overestimated by approximately 20\%.  

\section{Implications for the local value of Hubble Constant}

The multiplicative scaling of the parallax predicted from Cepheid
photometry is formally equivalent to a change in the zero point of the
calibration of the Leavitt Law; from the analysis in R16, this is
equivalent to a change in the estimated value of $ H_0 $.  The R16
value of $ 73.2 \hunit $ corresponds of course to a multiplicative
factor of 1.0, which is well within the 1-$\sigma$ range.  The
best-fit factor of $ \tgasmultfactor \pm \tgasmulterror $ ($
\pm \tgasmulterrorunity $ without rescaling the errors) would produce a
``recalibration'' to $H_0$=73.0 km s$^{-1}$ Mpc$^{-1}$ if interpreted
as such.  The \citet{planck16} value of $ 66.9 \hunit $, based on
Planck CMB data with {\lcdm}, corresponds to a factor of 0.91, which
would produce a change in total $ \chi^2 $ of 8.1 with the nominal
errors, or of 12.6 if the errors are scaled by a factor 0.8, and thus
a tension at the 2.5--3.5 $\sigma$ level, respectively, with the DR1
parallaxes.  However, it is difficult to determine whether there are
additional systematic uncertainties globally affecting this analysis.
Specifically, we do not know if we can reasonably combine so many low
SNR measurements to produce one, high SNR measurement without penalty
and thus the present result is tentative.

\begin{figure}[ht]
\includegraphics[height=\columnwidth,angle=90,bb=54 380 558 738] {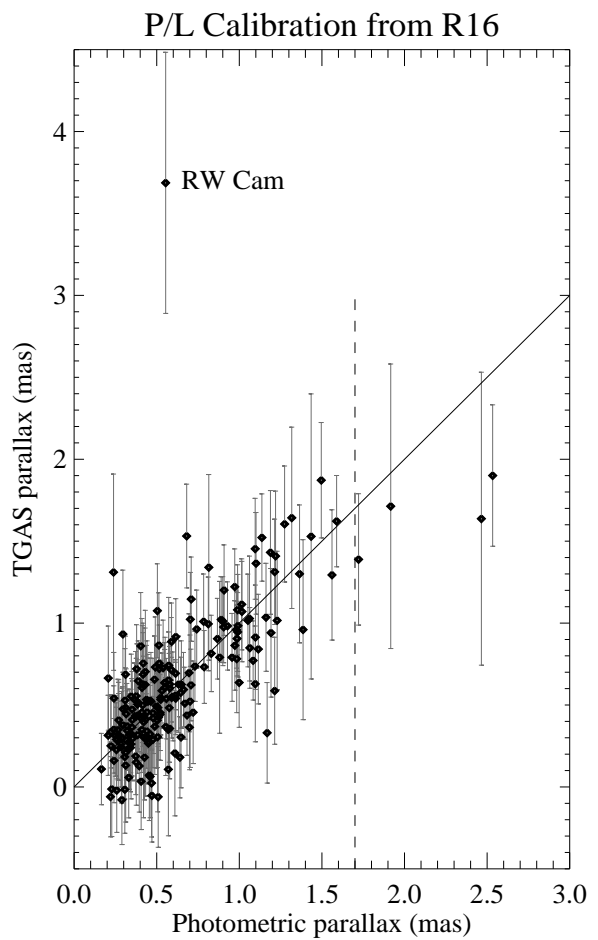}
\caption {\label{fig:tgas_phot} TGAS parallaxes vs.~photometric
parallax estimates.  The error bars are the formal errors as reported
in DR1.  Formal uncertainties in photometric parallax estimates are
small, about 5\%.  As noted in the text, the outlier RW~Cam is likely
to have a bright companion.  Note the apparent systematic offset for
stars with large photometric parallaxes; such stars also have larger
nominal errors.}

\end{figure}

\begin{figure}[ht]
\includegraphics[height=\columnwidth,angle=90] {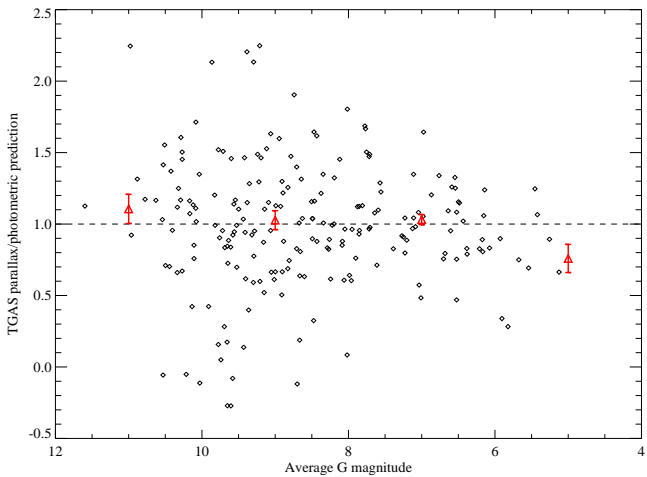}
\caption {\label{fig:tgas_ratio} Ratio between TGAS and photometric
parallaxes as a function of average G magnitude as given in the DR1
data.  The red symbols show the mean and the error in the mean
(based on the actual dispersion, not the nominal errors) for bins
2 mag wide.  Note that three stars
exceed the range of the Y axis and are not shown, but are included in
the averages.  This comparison suggests that TGAS parallaxes for very
bright stars ($ G < 6 $) may be systematically underestimated.}
\end{figure}

\begin{figure}[ht]
\includegraphics[height=\columnwidth,angle=90] {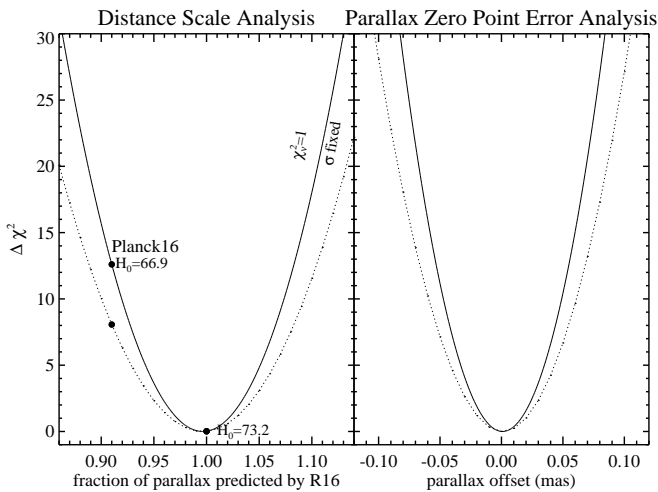}
\caption {\label{fig:systematics} Comparing Gaia and predicted
parallaxes to test either the distance scale or the parallax zero
point.  Shown is the change in $ \chi^2 $ caused by a multiplicative
term in the photometric parallaxes (left panel) or an additive 
term in DR1 parallaxes (right panel).  In each panel, the dashed
curve uses the nominal errors, while the solid curve is for errors
scaled by a factor 0.80, as suggested in
order to obtain $ \chi^2 = 1 $ per degree of freedom.  See text for
more discussion.}
\end{figure}

\section{Discussion}
   
The Gaia mission appears to be off to a tremendous start, and there is
little doubt that the full mission will produce results of great
import for cosmology.  While individual parallaxes have been measured
for a small number of Cepheids at better precision than Gaia DR1, the
new and exciting feature of DR1 is the angular breadth of the
measurements, providing for $\sim$ 300 $\mu$as precision for hundreds
of classical Cepheids.  Indeed, the comparison of the parallaxes
predicted for $ \sim 200 $ of these Cepheids suggests that the Gaia
DR1 uncertainties may have been conservatively {\it overestimated}.
The dispersion of $V,I$ and $H$-band based Wesenheit magnitudes and
log periods has been observed to be 0.08 magnitudes for over 1000
objects in the LMC \citep{macri15}, which would result in random,
individual errors of just $\sim 20 $ \muas.  The fact that the
$\chi^2_{\nu}=0.63 $ ($N=202$) likely results from a 20\% overestimate
of Gaia errors, as the errors in the predictions are too small to have
any impact.

The sample mean of the DR1 parallaxes for the Cepheids we consider has
a nominal error of $ \sim 20 \muas $.  {\it IF} there are no
systematic errors at this level in the DR1 measurements, then these
Cepheids would produce an independent calibration of the Cepheid
distance scale with an uncertainty of 3.1\% (original errors) to 2.5\%
(80\% errors), competitive with the best geometric calibrations from
NGC 4258 (2.6\%), previous MW Cepheid parallaxes (2.1\%) and detached
eclipsing binaries in the LMC (2.0\%) (R16).  The factor of twenty or
more improvement expected for Gaia parallaxes by mission end will push
the uncertainty due to geometric distance calibration well below 1\%,
assuming systematics can be kept under comparable control.
   
However, the geometric calibration of Cepheids, central to measuring
the Hubble constant, depends not on just the mean parallax precision of
the sample but also on the ability to compare them photometrically to
their cousins in distant galaxies.  The photometry of extragalactic
Cepheids in SN~Ia hosts can only be measured at present in space with
{\HST} and has been achieved most extensively with WFC3.  On the
other hand, the photometry of the Gaia DR1 Cepheid sample analyzed
here was obtained from various ground-based sources.  Due to the high
foreground extinction of the Milky Way fields and in external
galaxies, the use of near-infrared magnitudes and colors is especially
important.  Ground-based NIR filter systems are based on natural (and
nightly changing) breaks in water and OH emission and do not well
match the space based system.  This produces systematic uncertainties
at the level of approximately 0.02 mag, including the relative
uncertainties in NIR zeropoints and dereddened colors (R16).  These
uncertainties are currently below the precision of the geometric
calibration of the distance scale, but will be well above the
precision that can be achieved with Gaia full-mission results.

One of the best ways to mitigate the systematic error resulting from
comparing ground and space-based Cepheid photometry is to observe the
MW Cepheids with {\HST}'s WFC3.  We have undertaken a series of HST
programs to measure the magnitudes of $\sim$ 50 MW Cepheids with
relatively low extinction and we have employed rapid spatial scanning
with {\HST} to avoid the saturation which would occur with direct
imaging of such bright stars.  In the future, the combination of these
50 parallaxes from Gaia and their {\HST} photometry in F555W, F814W, and
F160W should produce a complete and effective calibration of
extragalactic Cepheids with a mean error of $\sim$ 0.5 \%, and an
anchor for a 1\% determination of the Hubble constant.

We have chosen to use the Gaia DR1 parallaxes as a test rather than an
augmentation of the current calibration of $H_0$ by R16 to avoid the
complication of Lutz-Kelker type bias corrections that would be large
and necessary for parallax measurements with mean SNR $ \sim 2 $,
and in recognition that these parallaxes are expected to dramatically
improve in the near term (thus reducing the need for such corrections
as well).

This is the start of an exciting phase of measurement and perhaps
discovery in the long-lived field of parallax measurement, with the
best yet to come.

\begin{acknowledgements}
This work has made use of data from the European Space Agency (ESA)
mission {\it Gaia} (\url{http://www.cosmos.esa.int/gaia}), processed
by the {\it Gaia} Data Processing and Analysis Consortium (DPAC,
\url{http://www.cosmos.esa.int/web/gaia/dpac/consortium}). Funding for
the DPAC has been provided by national institutions, in particular the
institutions participating in the {\it Gaia} Multilateral Agreement.

Support for this work was provided by NASA through program GO-13101
from the Space Telescope Science Institute, which is operated by AURA,
Inc., under NASA contract NAS 5-26555, and by the Agenzia Spaziale
Italiana (ASI) through ASI grant 2014-025-R.1.2015.
\end{acknowledgements}

\hide{Support for this work was provided by NASA through programs
\newtext{GO-12879} and GO-13101 from the Space Telescope Science
Institute, which is operated by AURA, Inc., under NASA contract NAS
5-26555.  A.V.F.'s group at UC Berkeley is also grateful for financial
assistance from NSF grant AST-1211916, the TABASGO Foundation, and the
Christopher R. Redlich Fund.  R.I.A. acknowledges funding from the
Swiss National Science Foundation as a postdoctoral fellow.  S.C. and
A.G.R.~gratefully acknowledge support by the Munich Institute for
Astro- and Particle Physics (MIAPP) of the DFG cluster of excellence
``Origin and Structure of the Universe.''  Research at Lick
Observatory is partially supported by a generous gift from Google.}

\hide{This research is based primarily on observations with the NASA/ESA
{\it Hubble Space Telescope}, obtained at the Space Telescope Science
Institute, which is operated by AURA, Inc., under NASA contract NAS
5-26555.}

\hide{This publication makes use of data products from the {\it
Wide-field Infrared Survey Explorer (WISE)}, which is a joint project
of the University of California, Los Angeles, and the Jet Propulsion
Laboratory/California Institute of Technology, funded by NASA.  It has
also made use of the SIMBAD database, operated at CDS, Strasbourg,
France.}

\bibliographystyle{aa}
\bibliography{mybibfile}

\end{document}